# A Physics-Constrained Deep Learning Model for Simulating Multiphase Flow in 3D Heterogeneous Porous Media

Bicheng Yan*, Dylan Robert Harp, Bailian Chen, Rajesh Pawar

Earth and Environmental Sciences, Los Alamos National Laboratory

*Corresponding author

Email: bichengyan@lanl.gov (B. Yan); dharp@lanl.gov (D.R. Harp); bailianchen@lanl.gov (B. Chen); rajesh@lanl.gov (R.J. Pawar).

**Abstract**

In this work, an efficient physics-constrained deep learning model is developed for solving multiphase flow in 3D heterogeneous porous media. The model fully leverages the spatial topology predictive capability of convolutional neural networks, and is coupled with an efficient continuity-based smoother to predict flow responses that need spatial continuity. Furthermore, the transient regions are penalized to steer the training process such that the model can accurately capture flow in these regions. The model takes inputs including properties of porous media, fluid properties and well controls, and predicts the temporal-spatial evolution of the state variables (pressure and saturation). While maintaining the continuity of fluid flow, the 3D spatial domain is decomposed into 2D images for reducing training cost, and the decomposition results in an increased number of training data samples and better training efficiency. Additionally, a surrogate model is separately constructed as a postprocessor to calculate well flow rate based on the predictions of state variables from the deep learning model. We use the example of $CO_2$ injection into saline aquifers, and apply the physics-constrained deep learning model that is trained from physics-based simulation data and emulates the physics process. The model performs prediction with a speedup of ~1400 times compared to physics-based simulations, and the average temporal errors of predicted pressure and saturation plumes are 0.27% and 0.099% respectively. Furthermore, water production rate is efficiently predicted by a surrogate model for well flow rate, with a mean error less than 5%. Therefore, with its unique scheme to cope with the fidelity in fluid flow in porous media, the physics-constrained deep learning model can become an efficient predictive model for computationally demanding inverse problems or other coupled processes.

## 1. Introduction

Multiphase fluid flow in porous media coupled with thermal, mechanical and chemical processes drives the overall performance of multiple subsurface storage and energy extraction applications such as geological $CO_2$ sequestration (Chen et al., 2018; Jiang et al., 2019), hydrocarbon recovery (Aziz and Settari, 1979) and geothermal recovery (Ezekiel et al., 2020; Fulignati et al., 2014). The governing equations used to describe multiphase fluid flow in porous media are highly nonlinear due to multi-scale heterogeneities (Efendiev et al., 2013; Yan et al, 2016), complex fluid thermodynamics (Michael et al., 2018), and coupled physics with thermal transport and geomechanics (Winterfeld and Wu, 2016). These governing equations can be accurately discretized by numerical methods (Chen et al., 2006), and for complex problems lead to large scale sparse linear systems that require parallel computing and scalable solvers (Fung et al., 2014; Wang and Killough, 2015). Reduced-order methods such as multiscale methods (Fish and Chen, 2004; Efendiev et al., 2013) and Proper Orthogonal Decomposition (Astrid 2004) have exhibited good performance for capturing the coherent spatial-temporal structures of state variables, such as pressure and saturation, by dimension reduction. These methods have been applied widely to problems involving multiphase fluid flow in porous media while achieving decent computational efficiency and accuracy (Markovinovic and Jansen, 2006).

Over the past several years there has been tremendous growth in applications of artificial intelligence, particularly deep learning (Goodfellow et al., 2016), to solve complex problems. With hardware advances, such as GPU machines, deep learning has been widely applied to numerous problems, including image recognition (He et al., 2016; Simonyan et al., 2014), speech recognition (Hinton et al., 2012; Serizel et al., 2016), robotics (Lillicrap et al., 2015; Won et al., 2020), medicine (Ronneberger et al., 2015; Senior et al., 2020), etc. The success of deep learning is due in part to its superior capacity to process high dimensional data (Georgious et al., 2020) and approximate various continuous functions (Csaji, 2001).

In the area of fluid flow in porous media, many efforts have been undertaken to enhance the capability to predict state variables (e.g., pressure and saturation) with deep neural networks. These include the application of deep feedforward networks and automatic differentiation (Baydin et al., 2017) to incorporate physics in loss functions during the training process (Sirignano et al., 2018; Raissi et al., 2019). Fuks and Tchelepi (2020) incorporated the 1-Dimensional (1D) Buckley-Leverett displacement equation into the loss function of a neural network, and investigated its predictive capability over different flux functions. Wang et al. (2020) proposed a Theory-guided Neural Network by imposing engineering and physics-based constraints on the neural network to predict single-phase flow in 2-Dimensional (2D) porous media. Harp et al. (2021) developed a physics-informed machine learning framework coupled with an automatically-differentiable physics model, and evaluated its feasibility in learning complex scenarios of subsurface pressure management. While the family of physics-informed machine learning approaches are suitable for predicting processes governed by physics with medium complexity, they may become too computationally expensive to solve problems with high nonlinearity; for example, multiphase flow in heterogeneous porous media where the physics kernels need to be efficiently differentiated.

Alternatively, image-based approaches have also been investigated to predict fluid flow in porous media. Zhong et al. (2019) developed a conditional deep convolutional generative adversarial network to map the nonlinearity in 2D multiphase flow in porous media, and achieved decent accuracy to predict the migration of a $CO_2$ plume in a heterogeneous aquifer. Mo et al. (2019) introduced a deep convolutional encoder-decoder network to predict dynamic multiphase flow in heterogeneous media based on a training strategy that combines a regression loss with a segmentation loss, and applied it for predicting $CO_2$ saturation and pressure snapshots. Klie and Florez (2020) proposed an Extended Dynamic Mode Decomposition (EDMD) integrated with a deep encoder to learn coupled fluid flow and geomechanics processes and validated the approach using shale gas production data from the field. Tang et al. (2020) developed a deep learning surrogate model, where a convolutional neural network regresses the nonlinear relationship between geological maps and fluid flow maps, and a recurrent network performs the temporal evolution. Their approach was able to history match and predict two-phase flow in 2D-transects of channelized reservoirs. With sparse connectivity between input and output, image-based approaches tend to have better efficiency for large scale models. Yet so far there is little work to investigate how the physics of fluid flow in porous media can be used to improve its predictive capability, especially in 3-Dimenionsal (3D) problems.

In this work, we introduce a physics-constrained deep learning model for predictions related to multiphase flow in 3D heterogeneous porous media. This model falls into the category of image-based approaches, since it takes full advantage of the spatial topology predictive capability of convolutional neural networks (CNN), specifically U-Net (Ronneberger et al., 2015). Our contributions to solve multiphase flow problems are fourfold. First, 3D reservoir domains are decomposed into manageable small 2D layer-wise images, and the vertical flow connectivity is effectively conserved following the concept of the two-point flux approximation. This effectively avoids using the expensive 3D convolutional operators and significantly increases the training data size. Second, as the propagation of pressure fronts in porous media is closely related to advective flux based on Darcy's law, an efficient continuity-based smoother is coupled with the U-Net as an additive smoother to ensure the spatial continuity that is missed by U-Net. Third, for the regions containing state-variable transients such as saturation fronts, two penalization schemes are proposed to regularize the training process such that the local resolution can be improved. Finally, a physics-based surrogate model is constructed to accurately predict well flow rate based on the state variables of pressure and saturation predicted by the deep learning models.

The paper is structured as follows. In Section 2, we briefly introduce the physics of multiphase flow in porous media, and illustrate the physics-constrained deep learning model and the surrogate model for predicting well flow rates in detail. In Section 3, the approach is tested on numerical experiments predicting $CO_2$ injection into a 3D heterogeneous saline aquifer, and the performance is comprehensively analyzed. In Section 4, we conclude the work with a few remarks.

## 2. Methodology for Physics-Constrained Deep Learning Model and Surrogate Model for Predicting Well Flow Rates

In this section, we first review the physics of multiphase fluid flow in porous media. Further, the physics-constrained deep learning model workflow is illustrated in detail, particularly focusing on the methodology to improve the efficiency and fidelity of the model. Finally, the surrogate model for predicting well flow rates is discussed.

### 2.1. Physics of Multiphase Fluid Flow in Porous Media

At a geologic $CO_2$ storage site where $CO_2$ is injected into saline aquifers, the fluid phases include a water-rich (liquid) phase and a $CO_2$-rich (supercritical) phase, and the fluid phase components include water and $CO_2$. The water-rich phase contains majority of water with dissolved $CO_2$ while the $CO_2$-rich phase contains majority of $CO_2$ with vaporized water. For our study we take into account dissolution of $CO_2$ in the water-rich phase but ignore vaporization of water in the $CO_2$-rich phase. Fluid transport in porous media is dominated by the advective flux based on Darcy's law. In a general unstructured compositional formulation (Yan, 2017), the governing equation of multiphase flow in porous media is the mass balance equation of each component expressed as

$$\frac{V_j}{\Delta t}\left(M_{i,j}^{n+1} - M_{i,j}^n\right) - \sum_c T_c \left(\frac{x_i \rho_w k_{rw}}{\mu_w}\Delta\Phi_w + \frac{y_i \rho_g k_{rg}}{\mu_g}\Delta\Phi_g\right) + \sum_s (x_i \rho_w q_w + y_j \rho_g q_g) = 0, \quad (1)$$

where the first term is the accumulation term for fluid storage, the second term is the advective flux based on the two-point flux approximation (Lude et al., 2007) and the third term is the source or sink term. Subscript $i$ represents the fluid component index ($i = water, CO_2$), subscript $j$ denotes the grid cell index, and $\alpha$ is the phase index ($\alpha$ = water-rich phase $w$, or $CO_2$-rich phase $g$). The three terms account for the mass of each component $i$ in each phase $\alpha$. $V_j$ is the volume of grid cell $j$; $\Delta t$ is the time interval from time step $n$ to $n+1$; $M_{i,j}^n$ is the moles of component $i$ per unit grid volume of grid cell $j$; $T_c$ is the transmissibility of connection $c$ with two connected grid cells; $x_i$ and $y_i$ are the component mole fractions of component $i$ in the water-rich phase and $CO_2$-rich phase, respectively; $\rho_\alpha$ is the phase molar density; $k_{r\alpha}$ is the phase relative permeability; $\mu_\alpha$ is the phase viscosity; $\Delta\Phi_\alpha$ is the phase potential difference of connection $c$ considering pressure, capillary pressure and gravity; $q_\alpha$ is the well volumetric flow rate of phase $\alpha$ calculated according to the Peacemean well model (Peaceman, 1983). For vertical wells, the well model is

$$q_\alpha = WI \frac{k_{r\alpha}}{\mu_\alpha}(p_{bh} - p_\alpha - \rho_\alpha g(z_{bh} - z)), \tag{2}$$

$$WI = \frac{2\pi \Delta z \sqrt{K_x K_y}}{\log\left(\frac{r_o}{r_w}\right)}, \tag{3}$$

where $WI$ is the well index, $p_{bh}$ is the well flowing bottom hole pressure, $p_\alpha$ is the phase pressure, $g$ is the acceleration due to gravity, $z_{bh}$ is the depth where $p_{bh}$ is measured, $z$ is the grid cell depth, $K_i$ is the rock intrinsic permeability in the $i$-th axis, $r_o$ is the effective wellbore radius and $r_w$ is the actual wellbore radius.

Several auxiliary relationships are used to constrain the mass balance equation, including balance between pore volume and fluid phase volume, capillary pressure relationship, and fluid thermodynamics equilibrium, and more details can be found in previous literatures (Chen et al., 2006; Li et al, 2019; Michael et al., 2018). Notice that the sum of fluid phase mole fractions equal to 1, as shown in Equation (4). Since the water vaporization in $CO_2$-rich phase is ignored in our study, $y_{water}$ is set to be 0.0.

$$\sum x_i = \sum y_i = 1, \text{ with } y_{water} = 0. \tag{4}$$

The mass balance equations combined with auxiliary relationships are solved iteratively to calculate the state variables pressure and saturation for each grid cell at each time step, and the secondary variables including well production rates are calculated using predicted state variables.

## 2.2. Physics-Constrained Deep Learning Model

Convolutional neural networks (CNN) have existed for decades (LeCun, 1989; Goodfellow, 2016) as a specialized deep learning approach to capture the spatial and temporal patterns in images through relevant filters. Through convolution operations, CNN has much sparser connectivity between input and output units than traditional neural networks, which can be treated as a regularized version of a multilayer perceptron (MLP). Similar to MLP, CNN can also be formulated as a recursive layered architecture, where each layer is a function of the layer prior to it, as shown in Equation (5),

$$F_l = \sigma(W_l \circ F_{l-1} + b_l), \tag{5}$$

where $F_l$ is the feature map of layer $l$, $\sigma$ denotes the nonlinear activation function, $W_l$ and $b_l$ are respectively the convolutional kernel and bias, and $\circ$ is the convolutional operator.

In modeling multiphase flow in porous media, the main task is to predict pressure and saturation (i.e., state variables) at each time step based on input feature variables such as permeability and porosity (i.e., static geological properties) and dynamic properties such as injection well controls. Generally, the output state variable and input feature variables are stored as 2D images or 3D volumes. For example, the permeability and porosity fields follow the distribution of geological facies. Therefore, we treat the prediction of state variables as an image-to-image translation task. Specifically, we leverage the spatial predictive capacity of a 2D CNN to efficiently perform the nonlinear regression (Ronneberger et al., 2015). However, in 3D reservoir simulations, most of the variables exist as 3D volumes instead of as 2D images. Therefore, effective 2D representations of the 3D volumes is required.

The physics of fluid flow in porous media requires that spatial continuity leads to changes in pressure and saturation at any location. In numerical simulations, the spatial connectivity in the Mass Balance Equation (1) is reflected by the advective flux term through the transmissibility $T_c$ that relates fluid flow between two connected grid cells as

$$T_c = \frac{T_1 T_2}{T_1 + T_2}, \tag{6}$$

$$T_i = \frac{K_i A}{d_i}, with\ i = 1, 2,\qquad(7)$$

where $T_1, T_2$ are the grid cell-based transmissibilities in connected grid cells 1 and 2, respectively, $K_i$ is the absolute permeability of grid cell $i$ along the connection direction, $A$ is the contact area of the two connected grid cells normal to the connection direction, and $d_i$ is the distance from the grid cell center to the contact area.

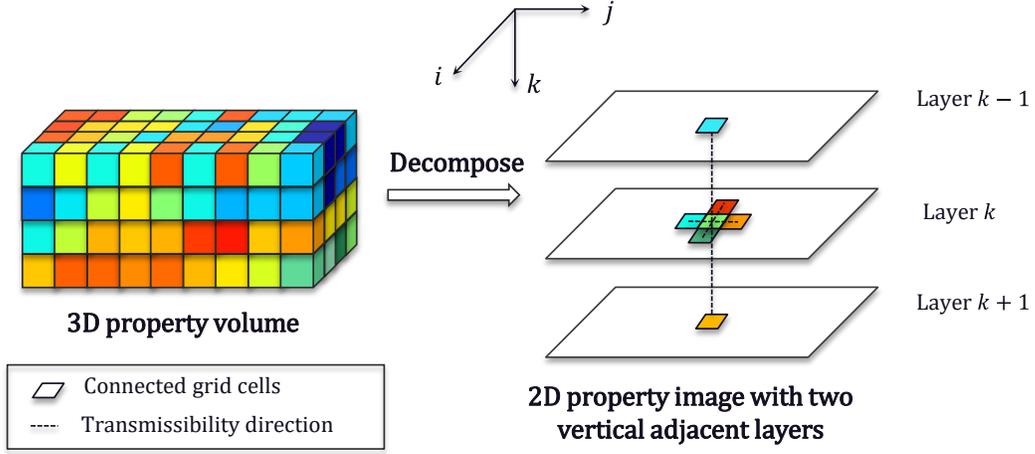

Fig. 1. Illustration of the decomposition of a 3D property volume into a 2D property image considering its connectivity with the upper and lower adjacent layers

Motivated by this, we slice the 3D volumes of permeability and porosity into horizontal layer-wise images (**Fig. 1**). Instead of predicting the state variables in the whole 3D volume simultaneously, we iteratively predict them for each layer. For a grid cell in layer $k$, , its transmissibility in the vertical direction is related to its adjacent connected grid cells in the layers $k-1$ and $k+1$ (**Fig. 1**). Therefore, layers $k-1$ and $k+1$ are included to capture the effects of vertical connectivity on layer $k$. For the top or bottom layers, since there is no adjacent upper or lower layer, respectively, we duplicate the layer for the missing adjacent layer. As a result of the decomposition into layers, the permeability and porosity fields become compatible with 2D convolutional operators. This approach also provides two additional benefits: (1) no need to use expensive 3D convolutional operators to process 3D volumes in a CNN (Milletari et al., 2016), especially if there are large number of layers in the 3D porous media domain; (2) increased number of data samples.

As opposed to the work by Tang et al. (2020), we do not use recurrent neural networks (RNN) to perform the temporal evolution of pressure and saturation. Instead, the time series data on well controls (e.g. production well bottom-hole pressure, injection well flow rate) are also transformed or broadcasted into 2D images, and fed into the CNN for temporal evolution. The feature list to predict the state variables is,

$$[p, S]^k = f(K^{k-1}, K^k, K^{k+1}, \phi^{k-1}, \phi^k, \phi^{k+1}, p_{bh}^k, Q_{inj}^k, t_i),\qquad(8)$$

where $p^k$ and $S^k$ are respectively the predicted images of pressure and saturation in layer $k$; $K^{k-1}, K^k, K^{k+1}$ are the permeability images of layers $k-1$, $k$ and $k+1$ respectively; $\phi^{k-1}, \phi^k, \phi^{k+1}$ are the porosity images of layers $k-1$, $k$ and $k+1$ respectively; $p_{bh}^k$ is the producer well bottom-hole pressure image in layer $k$; $Q_{inj}^k$ is the well injection rate in layer $k$ allocated through well index based on Equation (3); $t_i$ is a homogeneous image for the time step.

U-Net is a type of 2D CNN architecture (Ronneberger et al., 2015) with successive contracting and expansive steps. It is constructed from non-fully-connected CNN layers, and can efficiently capture spatial features and map complex nonlinear relationship between input and output images. Particularly, U-Net resolves the locality of images much better than other CNNs as it concatenates high resolution feature maps in the contracting steps with the upsampled ones in the expansive steps. Given that this architecture exhibits great performance in various image segmentation and regression tasks (Ronneberger et al., 2015; Zhong et al., 2019; Tang et al., 2020; Sun, 2020), we selected it as the basic convolutional network in our research.

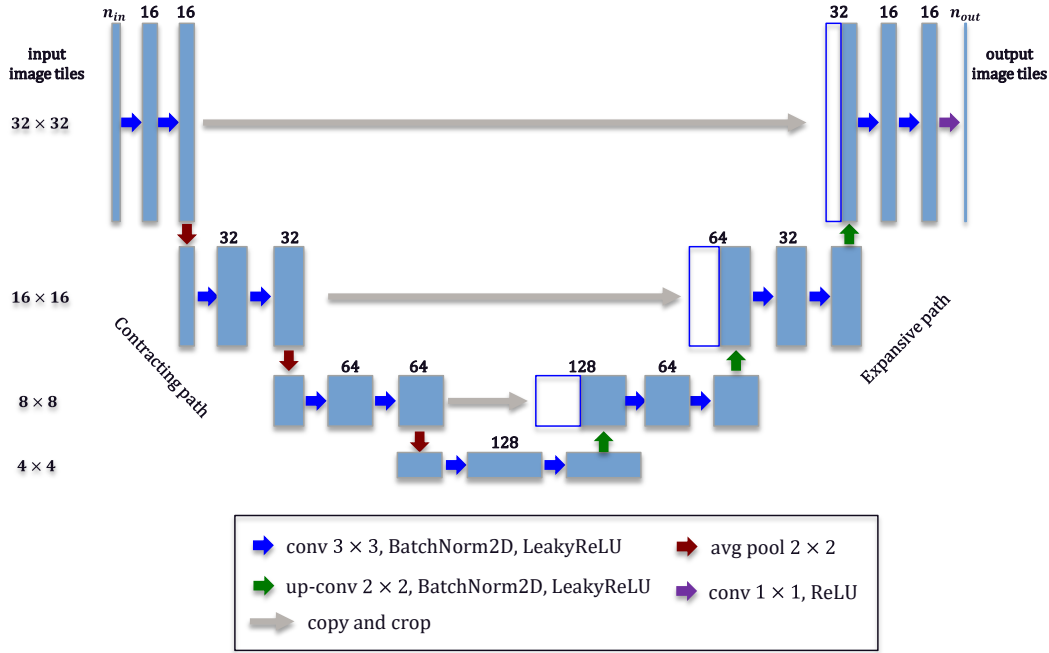

Fig. 2. Architecture of the U-Net in the work. The blue boxes represents multi-channel feature maps, the number of filters is labeled on top of each blue box, and the feature map size is shown along the left side. The empty boxes denote the cropped feature maps along the contracting path.

The architecture of the U-Net in this work is shown in **Fig. 2**, modified based on the original design of Ronneberger et al. (2015). In this work, the number of input channels or property maps (images) $n_{in}$ is set to 9 based on the feature list in Equation (8), but the number of output channels or property maps (images) $n_{out}$ can be either 1 or 2, depending on whether outputs of interest, namely pressure and saturation, are predicted separately or together. Here the number of channels represents the number of property maps or feature maps, and the number of filters in each convolution layer is equal to the number of output channels in the same layer.

In **Fig. 2**, the contracting path on the left side is a typical encoding network. At each contracting step, it performs downsampling operations including two convolution layers (blue arrow on the left side) and an average pooling layer (red arrow) before entering the next level. The number of output channels or filters in the path increases from 16 to 128. At each expansive step, it starts with an up-convolution layer (green arrows) that halves the number of channels, then concatenates with the high resolution feature maps at the corresponding contracting step (gray arrow), and ends with two convolutional layers (blue arrows on the right side). The number of output channels or filters decreases from 128 to 16 in this path. Through the output convolution layer (purple arrow), the number of channels is ultimately converted to $n_{out}$.

The activation functions used in the output convolution layer (purple arrow) is the rectified linear activation function (ReLU), defined as,

$$ReLU(x) = \begin{cases} x & if\ x \geq 0 \\ 0 & otherwise \end{cases}, \tag{9}$$

All other the activation functions used in other convolution layers (blue and green arrows) are leaky ReLU, defined as,

$$LeakyReLU(x) = \begin{cases} x & if\ x \geq 0 \\ 0.01x & otherwise \end{cases}. \tag{10}$$

Compared to the design of Ronneberger et al. (2015), the number of filters in this design is relatively small for reducing trainable parameters and improving the model efficiency. Additionally, based on a sensitivity analysis, average pooling is applied for downsampling instead of max pooling (Paszke et al., 2019). The design is generalized to take image tiles with arbitrary channel numbers of input and output images. This is helpful for us to take options to predict the state variables (pressure and saturation) separately or together.

For multiphase fluid flow in porous media, the propagation of a pressure front is much more rapid than that of a saturation front (Vasco, 2011). To better capture the spatial distribution of pressure, we designed a pressure smoother as a postprocessor to enhance the pressure prediction from the U-Net by considering the physics-based spatial continuity of pressure. Based on the physical constraints of local pressure continuity (Chen et al., 2013) and the numerical approximation of the mass balance equation, **Fig. 3** illustrates the finite difference stencil that computes pressure $p_{i,j}$ at a grid cell or node $(i, j)$ by taking the average of its neighboring grid cell pressures ($p_{i,j} = \frac{1}{4}(p_{i,j+1} + p_{i,j-1} + p_{i+1,j} + p_{i-1,j})$). This relationship ensures the spatial continuity in the predicted pressure field, which the U-Net is not constrained by.

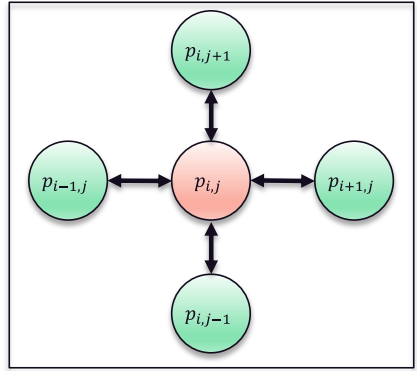

Fig. 3. Finite difference stencil for ensuring pressure continuity.

We construct a convolution kernel as a smoother which correlates pressure at a grid cell with pressures in the neighboring grid cells, namely continuity-based smoother. The kernel weight is initialized to have equal contributions of neighboring grid cells. Since the smoother is used to ensure the spatial connectivity with the reduced physics system, we treat it as an additive term to approximate the residual between the true solution and prediction from the U-Net as

$$p_{i,j} \cong p_{i,j}^{UNet} + F(p_{neighbors}^{UNet}), \tag{11}$$

where $p_{i,j}$ is the true solution of pressure at grid cell $(i,j)$, $p_{i,j}^{UNet}$ is the approximation of $p_{i,j}$ from the U-Net, $p_{neighbors}^{UNet}$ is the pressure prediction in the neighborhood of $p_{i,j}^{UNet}$, and $F$ is the smoother kernel operating on $p_{neighbors}^{UNet}$.

We illustrate the functionality of the smoother in **Fig. 4**. Specifically, it takes a 2D noisy image and scans through it by a square window size of $3 \times 3$ pixels, and enhances the spatial patterns in the output image. To ensure efficiency, the smoother can be treated as a single postprocessing convolution layer of the U-Net.

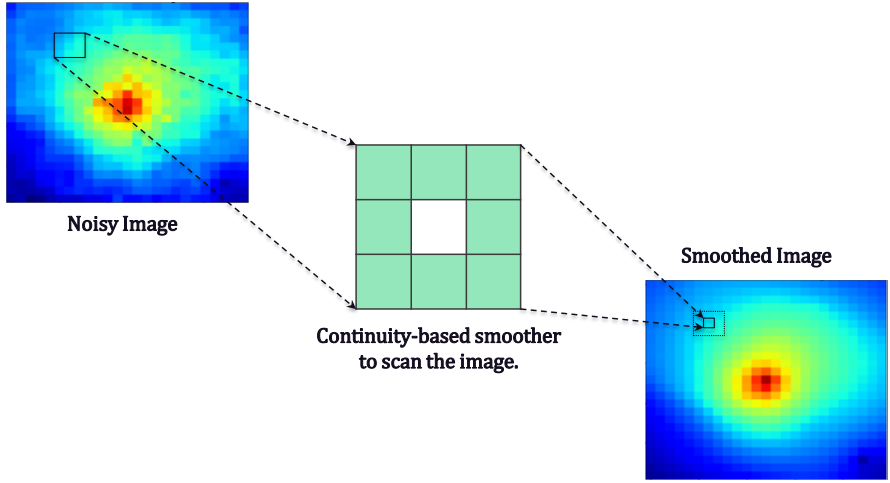

Fig. 4. A continuity-based smoother to ensure the spatial connectivity in the pressure field

Further, as there is particular interest in the transient regions with spatial gradients, e.g., pressure and saturation plume boundaries and well vicinities, the mismatches in these regions between ground truth and prediction of U-Net are used as penalties to weight the loss function of the U-Net, shown as Equation (12). A similar weighted loss was defined in the work of Tang et al. (2020) to improve the accuracy at the well locations. The goal is to find $\theta^*$ by solving the minimization problem, as shown in Equation (13),

$$\mathcal{L}(\theta) = \|Y_\Omega - \hat{Y}_\Omega\| + \sum_{\Omega_i} \lambda_i \|Y - \hat{Y}\|, \tag{12}$$

$$\theta^* = \underset{\theta}{argmin}\, \mathcal{L}(\theta), \tag{13}$$

where $\mathcal{L}$ is the loss function of the network; $\theta$ is a vector of trainable parameters (weights and biases) in the network; $Y$ is the ground truth of the state variable of pressure $p$ or saturation $S$; $\hat{Y}$ is the prediction corresponding to $Y$; $\Omega$ is the whole domain in each sample; $\Omega_i$ is a subset of $\Omega$, which represents regions of the model domain, including pressure plume, saturation plume and well vicinities; $\|\cdot\|$ is the root-mean-square-error operator.

As the pressure and saturation plumes evolve spatially and temporally, $\Omega_i$ in Equation (12) can be defined by one of the two options: (1) fixed $\Omega_i$ at all the time steps, (2) time-adaptive $\Omega_i$. In the first option a domain with fixed shape such as rectangular region surrounding the pressure and saturation plumes as well as around the production well (as shown in **Fig. 5**) is used. In the second option, a threshold value is used to filter the state variable field such that a time-adaptive zone can be generated as the state variable evolves with time as shown in **Fig. 6**. For both options, the region surrounding the fluid injection well is typically included within the regions selected based on the pressure and saturation plumes, which reduces the need to add additional penalty region for the injection well. On the other hand, a small rectangular region is always defined surrounding the production well in both options.

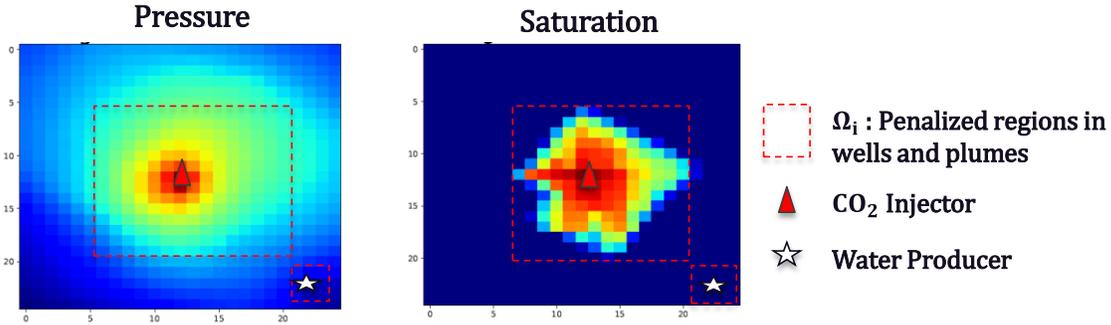

Fig. 5. Fixed penalty regions of pressure and saturation plumes and water production well vicinity

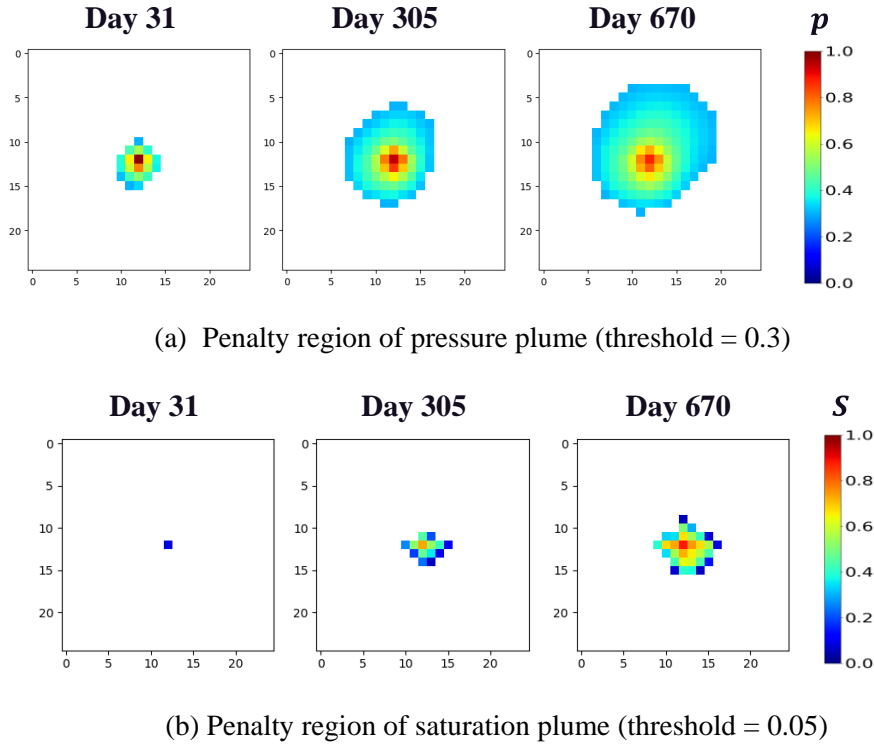

(a) Penalty region of pressure plume (threshold = 0.3)

(b) Penalty region of saturation plume (threshold = 0.05)

Fig. 6. Time-adaptive penalty regions of the pressure and saturation plumes. Colored cells are included in the region while white cells are excluded.

We implemented the U-Net and the smoother in the open source deep learning library, PyTorch (Paszke et al., 2019) and trained the U-Net on a single GPU (NVIDIA Quadro RTX 4000).

### 2.3 Surrogate Model for Predicting Well Flow Rate

The well flow rate is a key performance indicator for subsurface processes such as hydrocarbon recovery and geologic $CO_2$ sequestration. We focus on the well flow rate for the water production well which aims to manage reservoir pressures during geologic $CO_2$ sequestration. **Fig. 7** depicts a simple rectangular reservoir with a vertical well perforated in three different locations. During operation, reservoir fluid is extracted through the well perforations to surface processing facilities to increase available pore volume and decrease pressure.

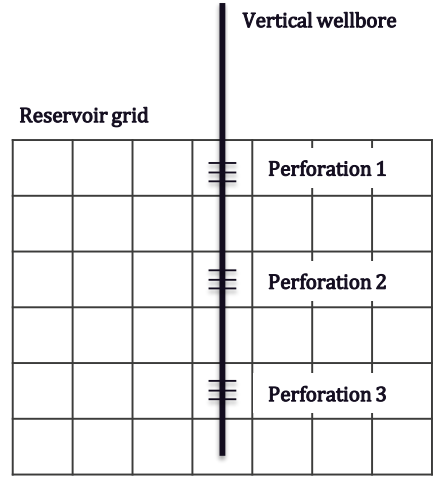

Fig. 7. A vertical well with 3 perforations in a reservoir.

Therefore, all of the perforations contribute to the volumetric flow rate of a producing wellbore as

$$Q_w^{std} = \frac{1}{\rho_w^{std}} \sum_{i=1}^{n_{perf}} \rho_{w,i} WI_i \frac{k_{rw,i}}{\mu_{w,i}} \left( p_{bh} - p_{w,i} - \rho_{w,i} g(z_{bh} - z_i) \right), \quad (14)$$

where $Q_w^{std}$ is the volumetric flow rate of water at standard conditions; $\rho_w^{std}$ is the water density at standard conditions, which is a constant; $n_{perf}$ is the number of perforations of the well; and other variables are defined for each perforation (denoted by subscript $i$) of the water-rich phase $w$, similar to Equation (2).

In Equation (14), water density $\rho_{w,i}$ and viscosity $\mu_{w,i}$ at each perforation are unknown variables if their empirical relationships with pressure are not provided. Therefore, with the gravity term $\rho_{w,i} g(z_{bh} - z_i)$ ignored, we define a nonlinear coefficient $\beta$ to consider the impact of $\rho_w^{std}$, $\rho_{w,i}$ and $\mu_{w,i}$. Equation (14) is rewritten as,

$$Q_w^{std} = \beta \sum_{i=1}^{n_{perf}} WI_i k_{rw,i} (p_{bh} - p_{w,i}), \quad (15)$$

All the terms on the right side of Equation (15) except $\beta$ can be calculated based on the static input parameters and the state variables of pressure and saturation that are predicted from the physics-constrained deep learning model. The relative permeability of water $k_{rw,i}$ can be calculated from the saturation and the well index $WI$ is function of permeability and well grid cell geometry.

As water density at standard conditions, $\rho_w^{std}$, is a constant value, and water density $\rho_{w,i}$ and viscosity $\mu_{w,i}$ at each perforation are functions of the grid cell pressure $p_{w,i}$ associated with the well perforations, then the nonlinear coefficient $\beta$ is a function of $p_{w,i}$ as well. **Fig. 8** presents the relationship of $\beta$ with $p_{w,i}$ at the three grid cells associated with well perforations based on our physics simulation data in this work.

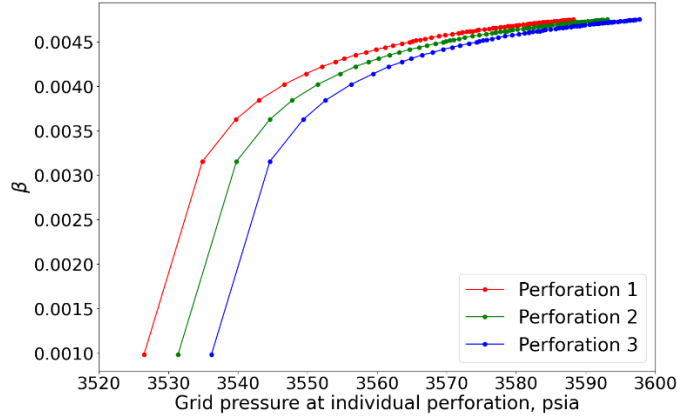

Fig. 8. Relationship between $\beta$ and grid cell pressure $p_{w,i}$ associated with well perforations

The relationship between $p_{w,i}$ and $\beta$ in the curves in **Fig. 8** can be easily approximated. Spefically we construct the surrogate model to predict $\beta$ using grid cell pressures $p_{w,i}$ by the gradient boost regression algorithm (Pedregosa et al., 2011), and the optimal number of boosting stages (estimators) in the regressor is 200 after hyper-parameter tuning. The surrogate model can be efficiently trained using CPU (Intel(R) Xeon(R) W-2125 CPU @ 4.00GHz) in 0.26 seconds, with a $R^2$ score ~98.4%. With $\beta$ predicted, $Q_w^{std}$ can be directly obtained through Equation (15).

### 3. Results

### 3.1 Heterogeneous 3D Reservoir Model

In this section, a physics-based model to simulate $CO_2$ injection into a 3D heterogeneous saline aquifer is introduced. The model is used to generate synthetic data to train and evaluate machine learning models discussed earlier. The physics-based model is developed using a commercial numerical reservoir simulator, GEM, by the Computer Modeling Group (CMG) (CMG, 2020). The reservoir is discretized using a uniform Cartesian grid, with $25 \times 25 \times 3$ grid cells in the $x, y, z$ directions, respectively. The grid cell size is $300 \times 300 \times 11.1\ ft^3$ and is constant throughout the domain. The permeability and porosity are correlated, and assumed to be spatially heterogeneous and uncertain. Multiple equiprobable realizations of heterogeneous permeability and porosity fields are generated using a geostatistical simulation approach and three realizations at P10, P50 & P90 (**Fig. 9**) are used.

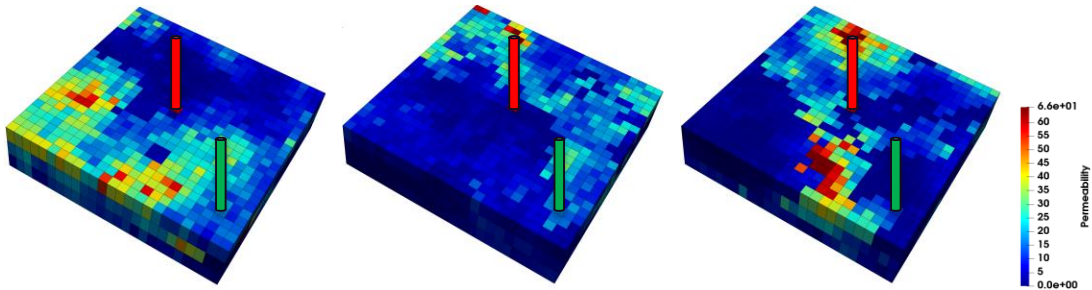

(a) 3 permeability realizations: left – P10; middle – P50; right – P90.

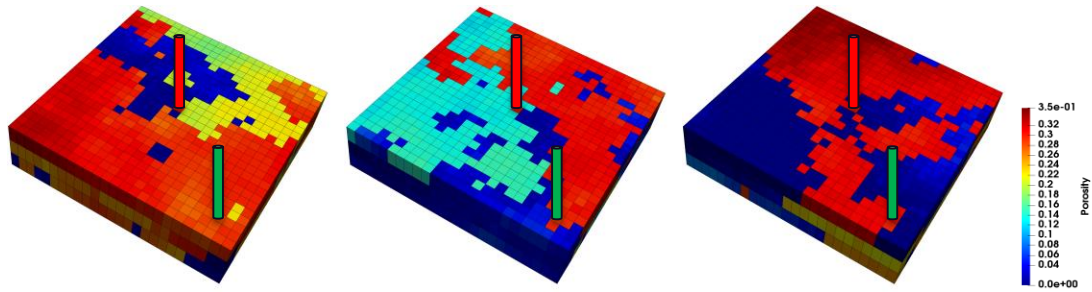

(b) 3 porosity realizations: left – P10; middle – P50; right – P90.

Fig. 9. Heterogeneous distributions of permeability and porosity in 3D reservoir model. Red cylinder: $CO_2$ injection well; green cylinder: water production well.

$CO_2$ is injected through an injection well located at the center of the domain and a water production well, located at the lower right corner, is used to manage reservoir pressure. Both wells are perforated in the 3 reservoir layers. Initially, the reservoir is fully saturated with the water-rich phase. During simulation, $CO_2$ is gradually injected into the reservoir layers with a maximum bottom-hole pressure of 5000 $psia$ and an injection rate ranging from 1000 to 9000 $mscf/day$ (1000 standard cubic feet per day). Water is produced from the reservoir with a minimum bottom-hole pressure set at 3525 $psia$. The mobile phases in the reservoir include a water-rich phase and a $CO_2$-rich phase, and the phase relative permeability is calculated from a look-up table based on the relations in **Fig. 10**. The simulation time is 5.67 years and there are 68 time steps.

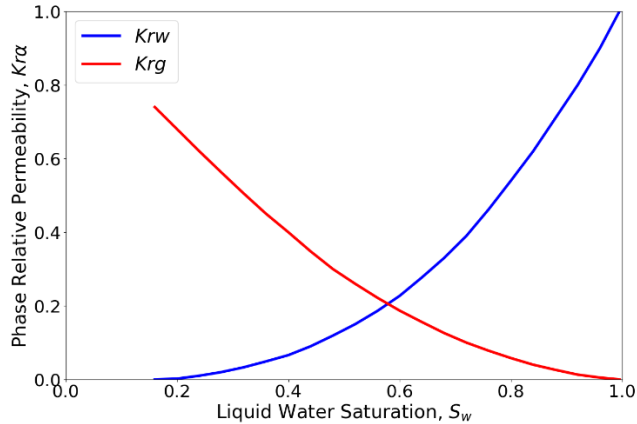

Fig. 10. Relative permeability curves for fluid with a water-rich ($w$) phase and a $CO_2$-rich ($g$) phase

A total of 27 simulation scenarios were performed using different combinations of the 3 correlated permeability-porosity fields and 9 $CO_2$ injection rates as shown in **Table 1.**

Table 1. Sampling parameters for the reservoir model

| Parameters | Number of Samples | Sample Ranges |
| --- | --- | --- |
| Permeability $K$ realizations | 3 | P10, P50, P90 |
| Porosity $\phi$ realizations | 3 | P10, P50, P90 |
| $CO_2$ injection rate | 9 | 1000 to 9000 |

We collected input and output data from the simulations, and split the simulation data into training, validation and testing sets based on the three parameters in **Table 1**, ensuring that each set of simulations has all the 3 different realizations of permeability and porosity fields. **Fig. 11** shows the data split with 3 validation cases (11%), 3 testing cases (11%) and 24 training cases (78%). The total number of available data samples is 5508 (27 cases × 68 time steps × 3 layers/case), since we treat each reservoir layer at a time step as an independent layer (**Fig. 1**). On the contrary, if the entire 3D data set at a time step is treated as a standalone sample, we will get only 1836 samples (27 cases × 68 time steps). The testing cases (Case 6, 13, 23 in **Fig. 11**) are not used in the training processes to have a fair comparison of the performance of the algorithms.

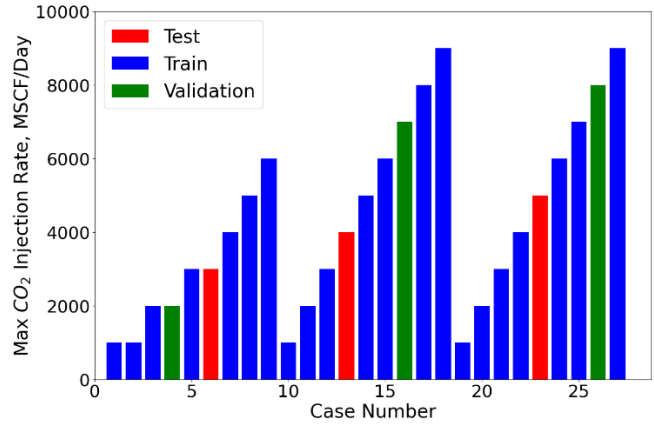

Fig. 11. $CO_2$ injection rates for the 27 simulation cases

### 3.2 Improvement from U-Net with Smoother

The smoother works as a postprocessor of the U-Net to predict state variable (pressure) that exhibits spatial patterns, and it needs to be synchronized and trained along with the U-Net (Equation (11). We use a U-Net with smoother to predict pressure and a separate U-Net without smoother to predict saturation. For comparison, we also evaluated a U-Net without smoother to predict pressure and saturation together, as the U-Net itself has the flexibility to handle various number of output images. These two options are shown in **Table 2**. The U-Net architectures of both options are based on **Fig. 2**. Since in both options a smoother is not used to predict saturation, they have similar saturation prediction accuracy as we demonstrate below.

Table 2. U-Net options to compare the performance

| Options | Specification |
|---|---|
| U-Net without smoother | Predict pressure and saturation by a single U-Net without smoother. |
| U-Net with smoother | Predict pressure by a U-Net with smoother, and predict saturation with a separate U-Net. |

For both options, the loss function is defined by Equation (12). Besides, two fixed regions (**Fig. 5**) are additionally used to regularize the loss, including $\Omega_1$ − a square region of size 15 × 15 grid cells centered at the $CO_2$ injection well to improve resolution of pressure and saturation plumes, and $\Omega_2$ − a square region of size 5 × 5 grid cells centered at the water production well. Through sensitivity analysis, the weights of $\Omega_1$ and $\Omega_2$ were respectively determined to be 0.3 and 0.05, which is consistent with the ratio of the reservoir model domain occupied by each penalty region. The training samples are split into small batches with 20 samples/batch, and the U-Net models are trained with an Adam optimizer with learning rate $10^{-4}$ and weight decay rate 0.01. The maximum number of epochs is 200 and the converging stagnant epochs is 20, such that the training process will stop when either 200 epochs are reached or 20 continuous stagnant epochs without better convergence have occurred.

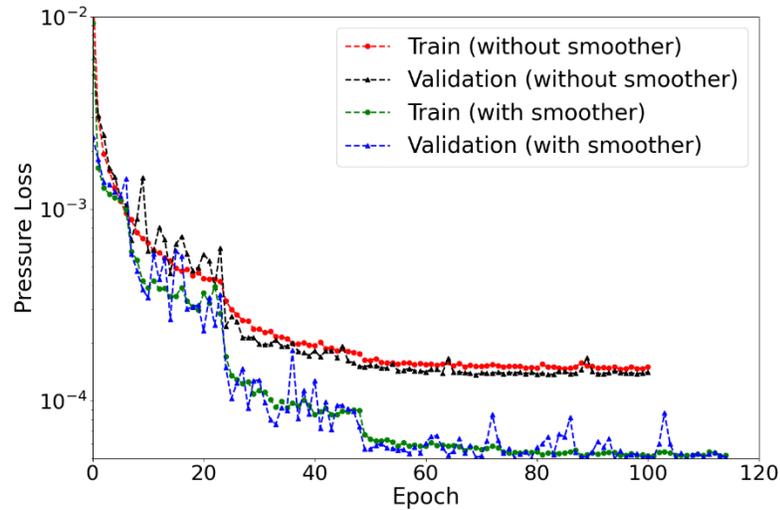

(a)

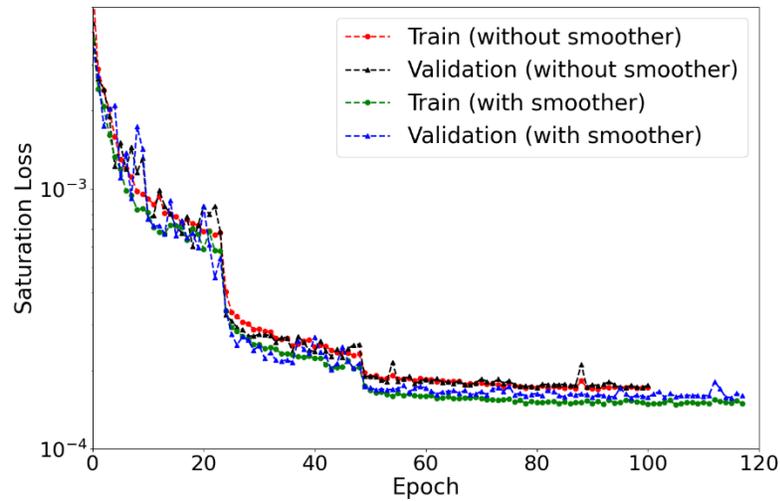

(b)

Fig. 12. Loss function comparison for options with and without smoother: (a) pressure, (b) saturation.

In **Fig. 12**, we compare the losses for the two options, with and without smoother, based on the training and validation data. Overall, the training and validation losses for all scenarios decrease in the same manner, which is an indication that the optimization did not overfit the model to the training data. The step-wise decrease observed in the loss curves during the training process, e.g. saturation loss around epoch 22, might be attributed to the optimization landscape of the loss function skewed by the penalty terms related to regions of $\Omega_1$ and $\Omega_2$. The ultimate pressure loss for the U-Net with smoother ($5.1 \times 10^{-5}$) is significantly lower than that from U-Net without smoother ($1.4 \times 10^{-4}$) due to the fact that the smoother effectively captures the spatial continuity in the predicted pressure field. While the use of smoother improves pressure predictions, it does not result in any significant benefit for saturation prediction as noted by the loss function plot on **Fig. 12(b)**. We provide further validatation of the improvement in pressure resolution at the pixel level through the testing data.

To measure the stability of temporal prediction of the deep learning models, the temporal error to predict the testing data is evaluated by Equation (16) based on Tang et al. (2020),

$$Y_{error}^t = \frac{1}{n_r n_g} \sum_{i=1}^{n_c} \sum_{j=1}^{n_g} \frac{\left\| Y_{i,j}^t - \hat{Y}_{i,j}^t \right\|}{\left(Y_{i,j}^t\right)_{max} - \left(Y_{i,j}^t\right)_{min}}, \tag{16}$$

where $t$ is time step; $Y$ is pressure or saturation; $n_r$ is the number of testing simulation runs; and $n_g$ is the number of grid cells for each reservoir simulation case.

**Fig. 13** shows the dynamic temporal error in predictions of pressure and saturation for both options (with and without smoother) based on the testing data. Over the entire injection operation, the temporal errors are highly stable, with mean pressure error less than 1% and mean saturation error ~0.1%. The temporal error in pressure for U-Net with smoother (mean: 0.27%) is ~1.2 times lower than that for U-Net without smoother (mean: 0.62%). The saturation temporal error with smoother (mean: 0.099%) is slightly better than without the smoother (mean: 0.115%). The dynamic trends in temporal prediction errors along with the overall loss performance (**Fig. 12**) gives confidence in the U-Net which was further tested with the testing cases.

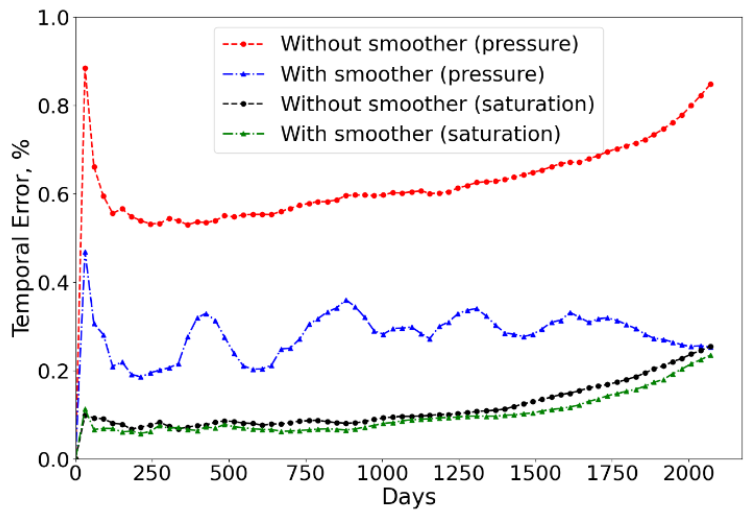

Fig. 13. Temporal error propagation for pressure and saturation.

Table 3. Basic parameters for 3 different testing cases.

| Testing Case | Permeability | Porosity | $CO_2$ injection rate, $mscf/day$ | Number of inactive grid cells |
|---|---|---|---|---|
| 6 | P50 | P50 | 3000 | 0 |
| 13 | P90 | P90 | 4000 | 13 |
| 23 | P10 | P10 | 5000 | 184 |

The parameters of the 3 testing cases are shown in **Table 3**. It's worth mentioning that the underlying reservoir model has a number of inactive grid cells with zero porosity and permeability values (Case 13 and Case 23) which increases the difficulty for U-Net without smoother. For conciseness, we present the predicted pressure and saturation snapshots at time step of 5.67 years for Cases 6 and Case 23, since the prediction is fairly stable with time (**Fig. 13**) and these two testing cases respectively have the lowest and highest number of inactive grid cells.

The pressure and saturation predictions for Case 6 and Case 23 are plotted in **Figs. 14** and **15** respectively. For Case 6 (**Fig. 14(a)**), the U-Net without smoother can roughly approximate the pressure plume shape, but it has substantial non-smoothness at the plume boundaries, with mean absolute error 13.35 $psia$. With the smoother, the mean absolute error of pressure decreases by 70% (3.926 $psia$), and the non-smoothness at pressure plume boundary is significantly reduced. The contrast between performances of U-Net with and without smoother is magnified in Case 23 where there are 184 inactive grid cells in the middle layer of the reservoir (**Fig. 15**). As seen in **Fig. 15(a)** the pressure predictions for U-Net without smoother have mean absolute error as high as 15.65 $psia$. The poor performance happens within both the discontinuous middle layer and its neighboring layers since pressure propagation follows spatial connectivity in the 3D space and the discontinuous permeability and porosity fields in this case bring significant challenge for the convolution operators in U-Net. However, coupling with the convolutional layer of the smoother as a postprocessor of the U-Net results in enhanced spatial connectivity in the pressure field by mimicking the functionality of the finite difference stencil (**Fig. 3**). As a result, the mean absolute pressure error of the U-Net with smoother decreases by 74% (4.032 $psia$) compared to the U-Net without smoother while maintaining higher fidelity even in the discontinuous portion of the grid cells in the middle layer. In **Figs. 14(b)** and **15(b)**, both approaches predict the saturation with high accuracy for the two testing cases and the difference between the predictions is quite small since the smoother is actually applied to the pressure but not the saturation. For all cases the saturation plume is significantly smaller than the pressure plume and mostly lies within the loss penalty area which is sufficient to steer the training of U-Net and better delineate the saturation plume shape with high resolution.

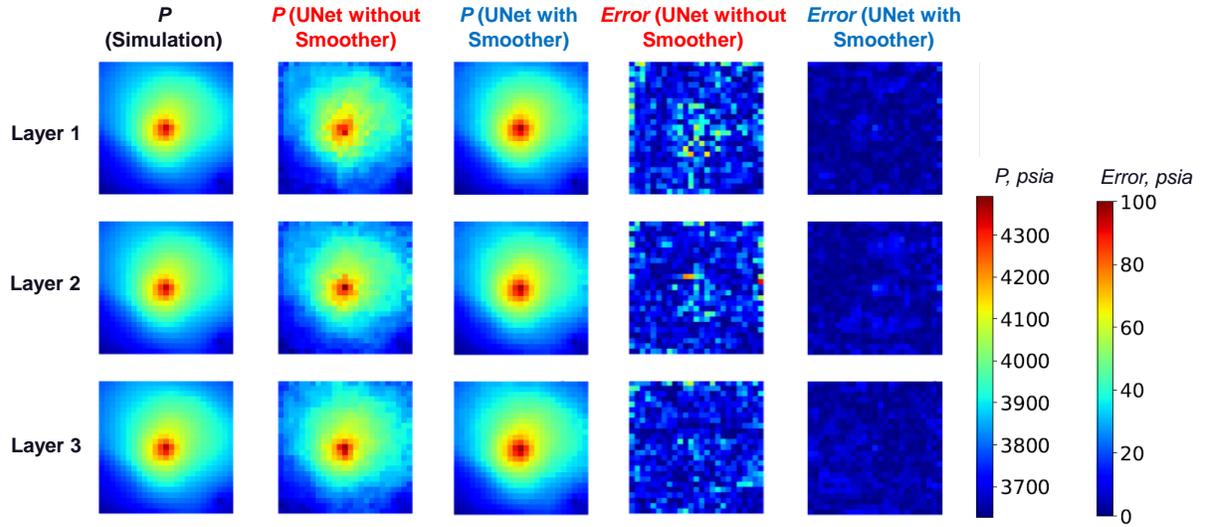

(a) Pressure. Mean absolute errors: 13.35 $psia$ for U-Net without smoother; 3.926 $psia$ for U-Net with smoother.

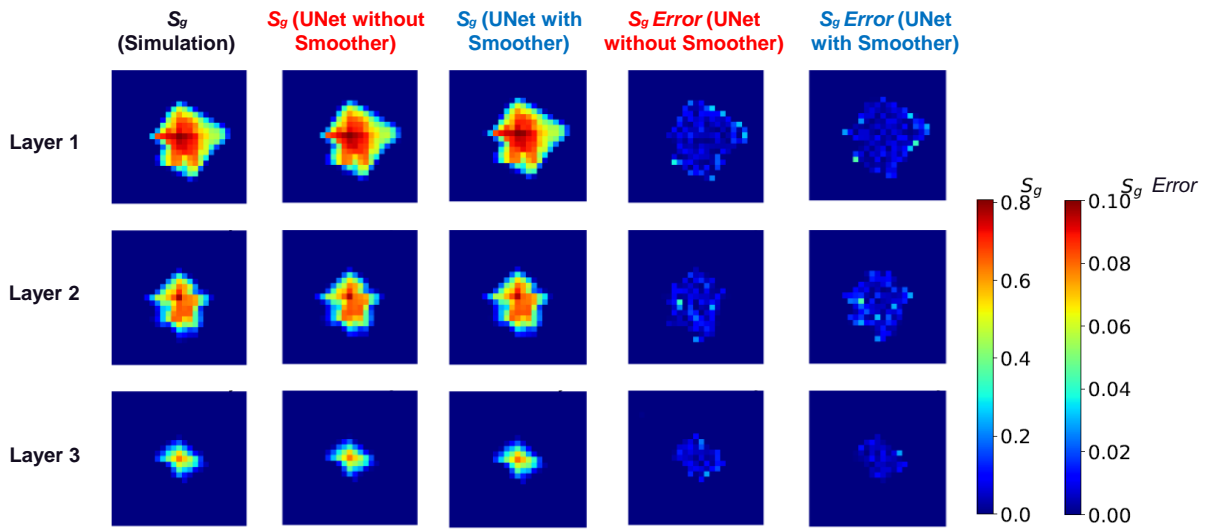

(b) Saturation. Mean absolute errors: 1.254e-3 for U-Net without smoother; 1.257e-3 for U-Net with smoother.

Fig. 14. Pressure and saturation prediction of Case 6 of all the 3 layers at 5.67 years.

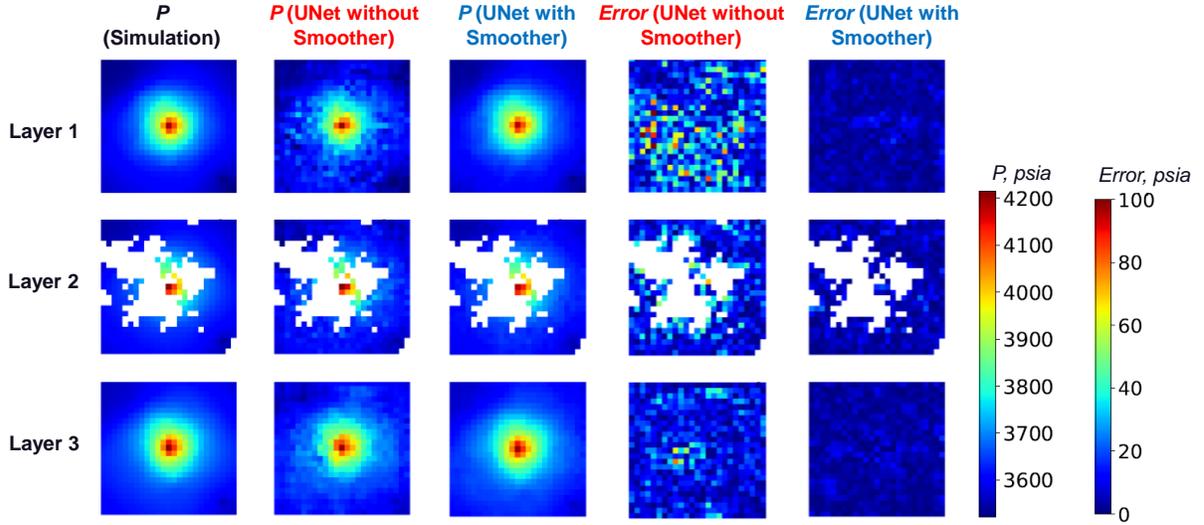

(a) Pressure. Mean absolute errors: 15.65 $psia$ for U-Net without smoother; 4.032 $psia$ for U-Net with smoother.

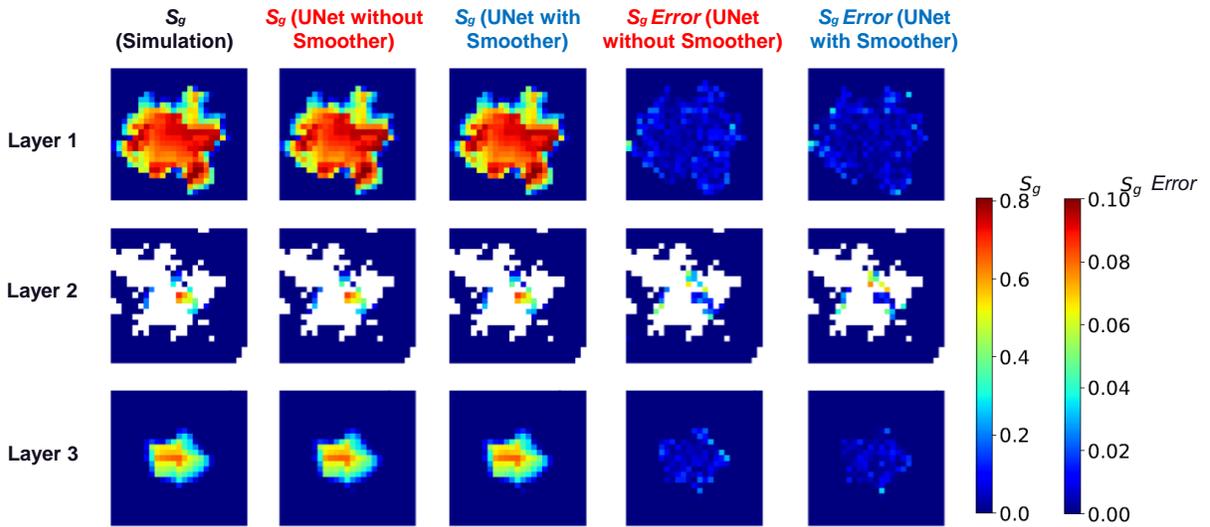

(b) Saturation. Mean absolute errors: 2.2e-3 for U-Net without smoother; 1.9e-3 for U-Net with smoother.

Fig. 15. Pressure and saturation prediction of Case 23 of all the 3 layers at 5.67 years.

The U-Net models provide predictions of pressure and saturation at the 3 grid cells perforated by the water production well. With the production well grid cell pressure, $p_{w,i}$, the surrogate model for well flow rate can be used to predict the nonlinear volumetric coefficient $\beta$ in Equation (15). The well index $WI$ is calculated from the production well grid-cell permeability and geometry, and the relative permeability of water is computed from predicted saturation through the water relative permeability curve (**Fig. 10**). With the above-mentioned terms the physics-based expression $\sum_{i=1}^{n_{perf}} WI_i k_{rw,i}(p_{bh} - p_{w,i})$ can be calculated, and ultimately the water volumetric rate, $Q_w^{std}$, is obtained from Equation (15). **Fig. 16** compares the water flow rate predicted by U-Net options with the ground truth solution from the simulation data, and **Table 4** summarizes the prediction errors. In all the three testing cases U-Net with smoother (average error 4.20%) performs better than U-Net without smoother (average error 7.90%). This is due to better prediction of pressure in the grid cell containing the water production well by the U-Net with smoother compared to the U-Net without smoother.

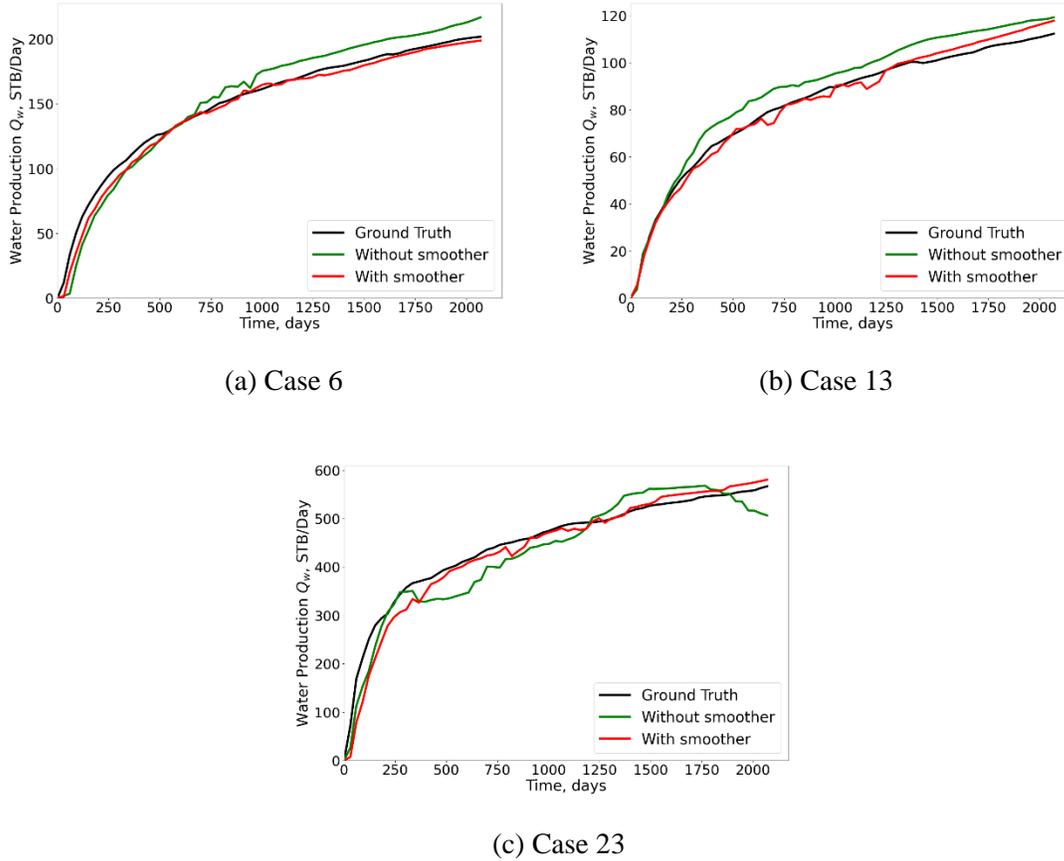

(a) Case 6

(b) Case 13

(c) Case 23

Fig. 16. Water rate prediction comparison. (a) Case 6; (b) Case 13; (c) Case 23.

Table 4. Water rate prediction error for U-Net with/without Smoother

| Cases | Well Rate Error | |
|---|---|---|
| | U-Net With smoother | U-Net Without Smoother |
| 6 | 7.61% | 3.39% |
| 13 | 8.15% | 3.37% |
| 23 | 7.93% | 5.85% |
| Average | 7.90% | 4.20% |

The training of the U-Net models is relatively efficient. Using a GPU (NVIDIA Quadro RTX 4000), it takes about 47 minutes to train the U-Net with smoother and 21 minutes to train the U-Net without smoother. The longer training time for the U-Net with the smoother approach is because two separate U-Net models for pressure and saturation are required to train, while only a single model for pressure and saturation is trained for the U-Net without smoother. However, the training time for the U-Net with smoother can be improved by parallel processing since the pressure and saturation models are independent and can be trained concurrently. **Table 5** lists the CPU cost of different modules in the physics-constrained deep learning model based on U-Net with smoother and the surrogate model for predicting well flow rate. It clearly shows that the trained model is much more efficient (1460 × speedup) than physics-based multiphase composition simulation. Even though it takes time to generate the training data with numerical reservoir simulations, the workflow becomes quite favorable for inverse problems that often require hundreds or even thousands of calls for forward simulations.

Table 5. CPU time cost of different modules in the workflow based on U-Net with smoother

| | Module Name | CPU Time, sec |
|---|---|---|
| Training | Data Cleansing & Normalization | 1.58 |
| | Well Flow Rate Surrogate Training | 0.26 |
| | U-Net Training | 2783 (47 min) |
| Prediction | U-Net Prediction* | 0.078 |
| | Well Flow Rate Prediction* | 0.015 |
| Physics-based Simulation Time (Averaged of 27 Cases)* | | 135.7 |

*CPU time for prediction is calculated based on 68 numerical time steps per case.

In **Fig. 17**, we also analyze the sensitivity of the number of samples per batch (batch size) on training cost and the corresponding accuracy for the U-Net with smoother. It shows that the training time scales well with the batch size on GPU resources and decreases as the batch size increases. This is interpreted to indicate that when training with larger batch sizes but similar numbers of epochs, the backpropagation of the deep neural network is calculated with lower frequency to update trainable parameters $\theta$ based on each batch. On the other hand, it is well known that larger batch size often leads to poorer generalization (Kandel and Castelli, 2020; Harp et el., 2021), and we also observe that larger batch size results in less accurate prediction, especially the pressure prediction. The results indicate a tradeoff between accuracy and training time and the optimal batch size in our study is 20 samples/batch.

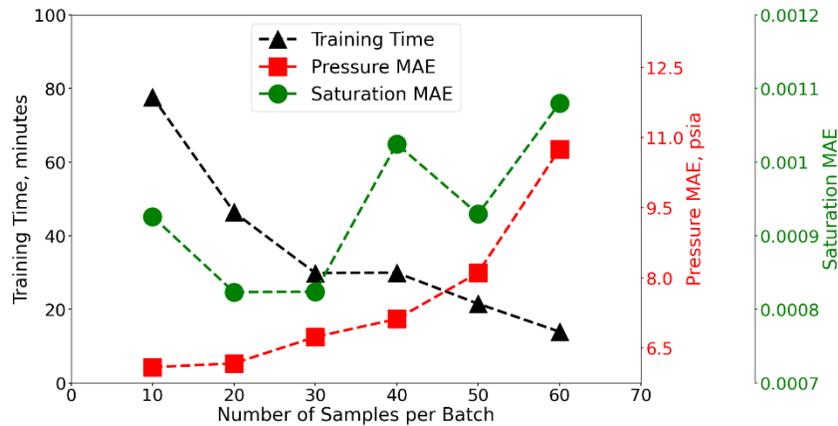

Fig. 17. Sensitivity analysis of sample number per batch (MAE: mean absolute error)

### 3.3 Performance of Adaptive Penalization Scheme

The time-adaptive penalty area scheme is a flexible approach to regularize the U-Net with smoother and improve the fidelity in the transient regions (**Fig. 6**). We analyze the scheme's performance using the same problem as above. As shown in **Fig. 6**, there are two time-adaptive penalty regions for pressure and saturation plumes surrounding the $CO_2$ injection well, and the threshold to filter the pressure plume penalty zone is 0.3 (normalized pressure), and that to filter the saturation plume is 0.05 (normalized saturation). Besides, there is a square zone of size $5 \times 5$ grid cells centered at the water production well. Consistent with previous models, the weights of the saturation and pressure plumes in this model are set to 0.3 and the weight of the region for the water production well is 0.05. Parameters related to training processes are set exactly the same as the models in sub-section 3.2.

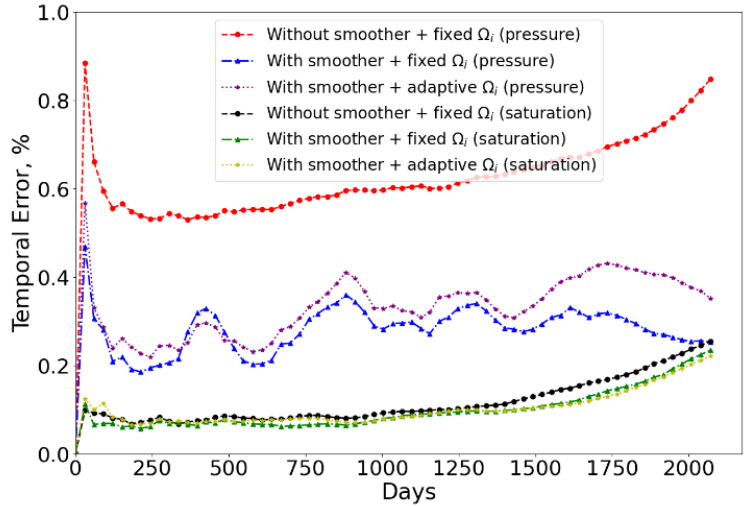

Fig. 18. Temporal error propagation for pressure and saturation based on different models

For conciseness, we analyze the temporal prediction of the U-Net with smoother and adaptive penalty region ($\Omega_i$) by comparison with the two models in sub-section 3.2. As shown in **Fig. 18**, with adaptive $\Omega_i$ in U-Net with smoother (purple line), the pressure error is quite stable and close to that with smoother and fixed $\Omega_i$ (blue line), and the slight difference of their errors is likely due to the fact that the fixed $\Omega_i$ does a better job considering the the maximum area of transient regions in this reservoir model, while the adaptive $\Omega_i$ may miss some of the boundary grid cells in the transient regions when filtering the domain by the threshold. As for saturation prediction, it is always stable and accurate for all the three models. Further, we present the prediction based on Case 23 with the most number of inactive grid cells. In **Fig. 19**, both the pressure and saturation predictions are compared with the ground truth from the physics-based simulation. The pressure can be predicted with high accuracy, and mean absolute error (MAE) is 4.301 $psia$, which is very close to the U-Net with smoother and fixed penalty scheme (MAE: 4.032 $psia$) (**Fig. 15 (b)**). Overall, the saturation is also quite accurate with MAE close to the other two models in **Fig. 15**. These results demonstrate that the adaptive penalty scheme can also be a good option to regularize convolutional network models. When the transient regions are in irregular shape and are spatially inter-connected, e.g., well interference in fluid flow in porous media, this option can be superior to penalty region with fixed $\Omega_i$.

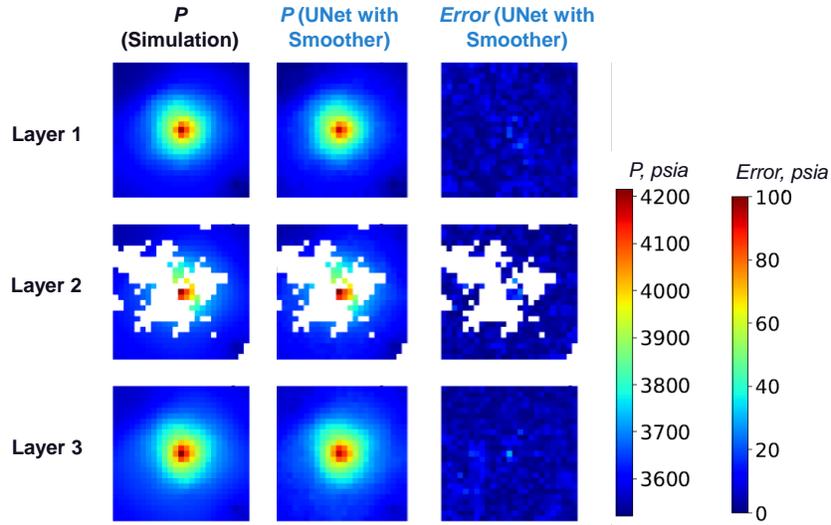

(a) Pressure. Mean absolute errors: 4.301 $psia$.

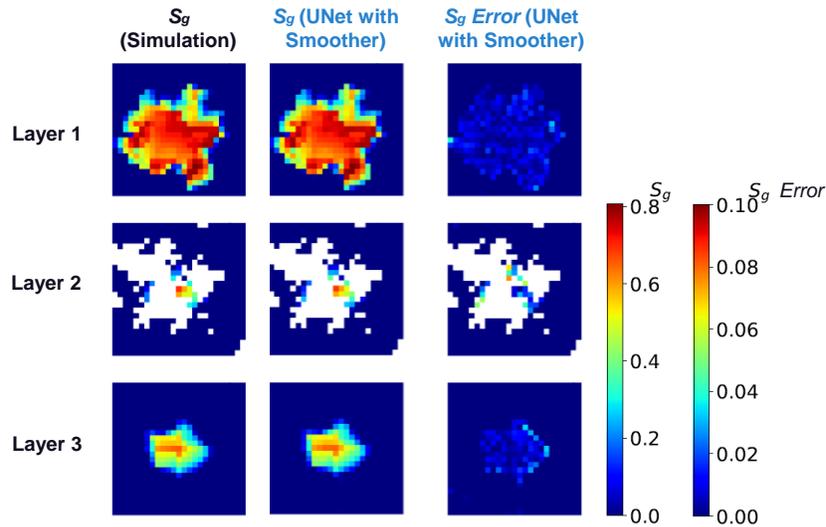

(a) Saturation. Mean absolute errors: 2.031e-3.

Fig. 19. Pressure and saturation prediction of Case 23 of all the 3 layers at 5.67 years based on U-Net with smoother and adaptive penalty scheme for plumes.

## 4. Discussion and Conclusion

In this study, a physics-constrained deep learning model to learn multiphase flow in 3D heterogeneous porous media is developed using a customized U-Net and a surrogate model for predicting well flow rate. The U-Nets are trained to predict the spatial and temporal state variables such as pressure and saturation, and the surrogate model is developed as a postprocessor that uses the U-Net predictions to calculate the well flow rate. In order to fit the data structures in efficient 2D convolutional operators in a U-Net, the original 3D spatial domain is decomposed into horizontal layer-wise images while conserving the vertical flow connectivity of neighboring layers which also leads to an increased number of data samples. To ensure the spatial continuity in pressure prediction, a convolutional layer smoother based on the simplified governing physics of fluid flow in porous media is designed to couple with U-Net. Additionally, a loss function penalty is added for transient flow regions to regularize the training of deep learning models, such that the predictive accuracy in these regions can be improved.

We demonstrated performance of the workflow by applying to the problem of $CO_2$ injection into a saline aquifer. We demonstrated that the smoother significantly improves the spatial pressure predictions even for challenging problems with large areas of inactive grid cells. Compared to the U-Net without the smoother, this approach can accurately predict pressure with temporal relative error as low as 0.27% (1.2 times lower than that without the smoother) and reduce the mean absolute error by more than 9 $psia$ (>70%). Since saturation and pressure are trained independently, the saturation plume can be predicted with high accuracy and stability without a smoother, with temporal relative error ~0.1% and mean absolute error ~1.0e-3. Given the smaller area of saturation plume compared to the pressure plume, the fixed and dynamic penalty loss term can effectively delineate the saturation plume during training and can provide high fidelity predictions. The U-Net with smoother is extremely computationally efficient with a 1460 times speedup compared to the physics-based multiphase compositional simulation. The U-Net presents superior scalability with batch size when training on GPU and an optimal batch size can be determined to balance predictive accuracy and efficiency. Finally, informed by the Peaceman well model, the surrogate model for well flow rate can predict water production rate with an average accuracy of 4.2%. This helps to avoid direct prediction of well flow rate with U-Net, as long as the state variables of pressure and saturation are accurately predicted.

## CRediT authorship contribution statement

**Bicheng Yan**: Study conceptualization, methodology development, software development, computational analysis, manuscript preparation and revision. **Dylan Robert Harp**: Supervision, study conceptualization, methodology development and manuscript revision. **Bailian Chen**: Study conceptualization, methodology development and manuscript revision. **Rajesh Pawar**: Funding support, supervision, study conceptualization, methodology development and manuscript revision.

## Acknowledgements

The authors acknowledge the financial support by US DOE's Fossil Energy Program Office through the project, Science-informed Machine Learning to Accelerate Real Time (SMART) Decisions in Subsurface Applications. Funding for SMART is managed by the National Energy Technology Laboratory (NETL). The authors also thank Dr. Seyyed A. Hosseini from University of Texas at Austin for providing the reservoir simulation data for $CO_2$ geological sequestration, and thank Dr. Diana Bacon from Pacific Northwest National Laboratory for providing parsing tools to process simulation data.